\documentclass[12pt]{iopart}
\pdfoutput=1
\usepackage{graphicx}
\usepackage[latin1]{inputenc}
\usepackage{amsfonts}
\usepackage{amssymb}
\usepackage{color}
\graphicspath{{./images/}}

\begin{document}
\title[The free lunch of a scale-free metabolism]{The free lunch of a scale-free metabolism}
\author{D.De Martino$^{1}$}
\address{$^1$ Institute of Science and Technology Austria (IST Austria), Am Campus 1, Klosterneuburg A-3400, Austria\\}
\begin{abstract}
In this work it is shown that scale free tails in metabolic flux distributions inferred from realistic large  models can be simply an artefact due to reactions involved in thermodynamically unfeasible cycles, that are unbounded by physical constraints and able to perform work without expenditure of free energy.  After correcting for thermodynamics,  the metabolic space scales  meaningfully with the physical limiting factors, acquiring in turn a richer  multimodal structure  potentially leading to symmetry breaking while optimizing  for objective functions. 
\end{abstract}

\section*{Introduction}
How physical constraints shape the space of feasible living beings?
Could we have a bacterium 1 meter long and/or  men as tall as buildings?
Typically natural systems are characterized by certain {\em scales}, e.g.
it is well known  from statistical physics that we can break a magnet and get two downto  a certain size, known as  correlation length\cite{cardy1996scaling}.
Scale-free systems however are present in nature, the proven example being equilibrium  thermodynamic systems at criticality, ruled by a beautiful and deep physical theory\cite{zinn2002quantum}. 
The wide and rich response to perturbations that these systems endow as well as  intriguing observations\cite{schneidman2006weak} have raised the quest for scale free processes and structures in biology\cite{mora2011biological} with interesting ongoing discussions, e.g. on biological neural networks structure and dynamics\cite{levina2007dynamical, mastromatteo2011criticality, tkavcik2015thermodynamics}.
A key biological process with deep physical roots ideally at the core of emergence of scales in living systems is metabolism, i.e. the fundamental set of reactions building up cell components and  transducing free energy. The energetic constraints of metabolism have been  considered to play a role in many important issues, from eucaryotes evolution  to aging, all the way down to the origin of life itself\cite{lane2015vital}.
Nowadays it is possible to keep track for a given organism the entire set of enzymes devoted to metabolic functions upto to reconstruct the entire chemical reaction network at genome scale level. The analysis of such complex systems is difficult, simplifying assumption are maded into a framework known as constraint based modelling  and/or flux balance analysis\cite{orth2010flux} with wide applications e.g. in metabolic engineering\cite{patil2004use}. Apart from practical applications are there general laws ruling the large scale organization of metabolic networks? 
Based on general inferences on these models,
it has been argued within the framework of  complex network theory\cite{barabasi2011network} 
that metabolic fluxes are globally organized in a {\em scale free} manner around a backbone of reactions carrying high-intensity fluxes\cite{almaas2004global}. 
While it is generally acknowledged the existence and importance of  central core pathways in metabolism, it is generally believed that such core has however {\em scales} that are  simply given by the physical limiting factors, being them extrinsic (resources avalability\cite{monod1949growth}) or intrinsic (maximum ribosome elongation rate\cite{bremer1996modulation}).   
Further, it is generally known  in the field of flux balance analysis that high intensity fluxes retrieved in models can be a spurious effect due to an erroneous handling of thermodynamics constraints. The way these constraints impact and shape the space of feasible states\cite{beard2002energy} is thus very important and attracted a lot of attention in recent years\cite{ataman2015heading}
upto the deeper clarification of its  theoretical basis from a mathematical\cite{reimers2014metabolic} and physical point of view\cite{polettini2014irreversible}. In this work it is showed that once thermodynamic constraints are implemented the long tail in flux intensity distribution  is truncated and the whole space scales meaningfully with the physical limiting factors, e.g. the availability of resources and/or the dilution rate.
On the other hand it is pointed out that the space is not anymore convex
thus acquiring a rich multimodal structure possibly leading to symmetry breaking upon optimizing for objective functions. After illustrating the background, we analyze in particular a genome scale reconstruction of the metabolism of the bacterium E. Coli and illustrate the mechanisms  in simple toy models.

\section*{Results}
\subsection*{Background.}
The assumption behind constraints based modelling is essentially to consider metabolism as a well mixed  chemical reaction network in the steady state. Concentration variations of   metabolites  are linearly related by mass balance to reactions fluxes through the matrix of stoichiometric coefficents 
\begin{equation}
\dot{c}_\mu = \sum_i S_{i \mu} f_i 
\end{equation}
Once a steady state is assumed and reaction fluxes are bounded by physical limits (reversibility, medium, kinetics) the reactions fluxes span a convex polytope
\begin{eqnarray}
\sum_i S_{i \mu} f_i =0 \\
f_i \in \left[ f_{i,min},f_{i,max}\right]
\end{eqnarray}
When reaction bounds are not known/provided, it is costumary to set them to an arbitrary  large number, e.g. $f_i \in \left[-10^4,10^4\right]$,  that, for a meaningful model, shall not influence the results.
In the most simple way thermodynamics is implemented by setting reaction reversibility, i.e. $f_{i}\geq 0$ for some reactions. On the other hand a more rigorous yet simple approach consists in postulating that  fluxes shall follow at least a free energy gradient\cite{deMartino12}, i.e. if $f_i \neq 0$, then $f_i \Delta G_i < 0$ where $\Delta G_i$ is the free energy drop of reaction $i$. Upon decomposing the reaction free energy in terms of chemical potentials we have a system of linear inequalities ($\xi_{i \mu} = - sign(f_i) S_{i \mu}$) 
\begin{equation}
\sum_\mu \xi_{i \mu} g_\mu >0  \quad \forall i 
\end{equation}  
whose feasibility is necessary for the thermodynamic soundness of the flux configuration. The feasibility of such a system is ruled by  the Gordan theorem by which the dual system
\begin{equation}
\sum_i \xi_{i \mu} k_i =0 \quad k_i\geq 0 \quad \forall \mu 
\end{equation}
shall have no non-trivial solution(s). It is thus seen that any solution of the latter system, that has the form of a closed cycle, it provides an unfeasible ``gauge invariance''\cite{DeMartino:2013p4115} for the involved fluxes such that, if $f_{i,0}$ is a feasible state and $k_i$ is such a solution, the line $f_i(L) = f_{i,0} + k_i s_i L$   is still inside the polytope  till $L$ reaches the (arbitrary)  bounds. 
From a physical point of view it corresponds to a ``perpetuum mobile'' from which we could get work for free.
A simple illustration is provided in the toy network depicted in Fig 1, where we see that the variable $x$ reaches unfeasibly high values without taking into account properly thermodynamics, whose implementation leads on the other hand to the non linear constraints $u x \geq 0$, $|x|\leq |u|$.
In essence thermodynamics forbid certain flux directions cutting in turn orthants within the feasible space that get decomposed in a connected star of convex polytopes, that is overall not convex anymore (see Fig 1), but whose scales now shall reflect the truly physical constraints of the given model. If we suppose that the flux $x$ is performing work we would have perpetual motion outside the thermodynamically feasible space (a ``realistic'' example for ATP production has been shown in \cite{deMartino12}). 

\begin{figure}[h]
  \fbox{\begin{minipage}[]{120pt}
    \includegraphics[height=100pt,width=120pt,angle=270]{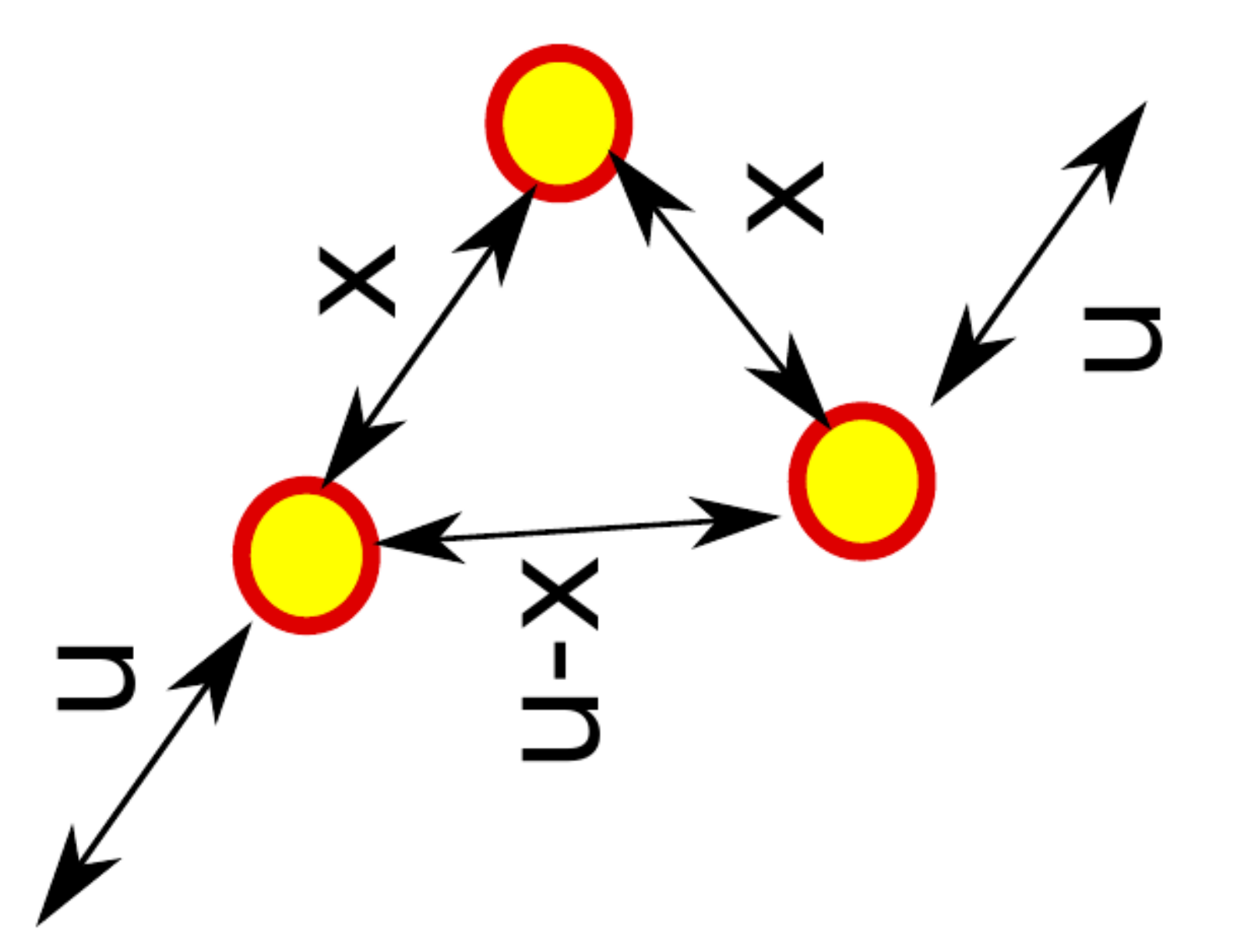}
  \end{minipage}}
  \hfill
  \fbox{\begin{minipage}[]{300pt}
    \includegraphics[height=150pt,width=300pt]{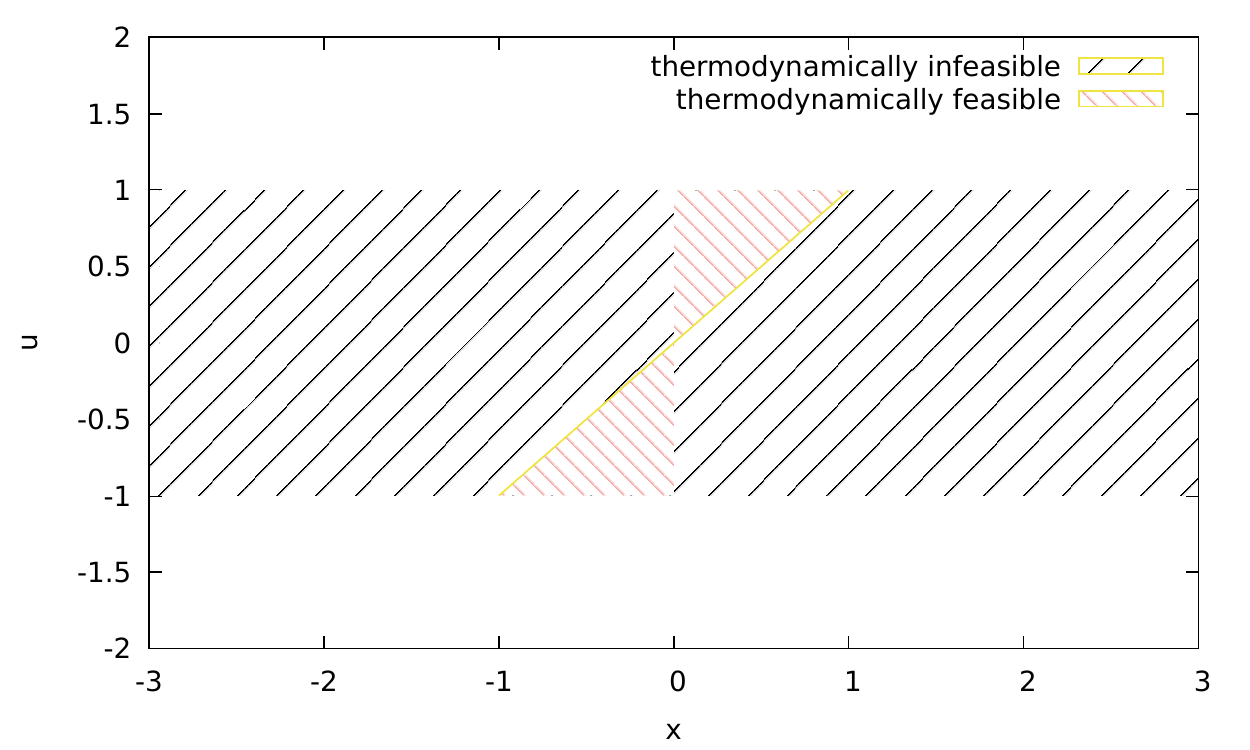}
  \end{minipage}}
  \caption{Left: toy network model illustraing the mechanism. Right: the plane $(x,u)$, in the case where $u\in[-1,1]$ and $x\in[-3,3]$. Thermodynamics constraints impose further $ux\geq 0$ and $|x|\leq |u|$. Notice that the thermodynamically feasible space is not convex, it scales with the bounds of $u$ and leads to bimodal distributions.}
\end{figure}

\subsection*{E. Coli genome scale model analysis.}
We consider here  the typical steady states of the genome scale model of the E Coli metabolic network iJR904\cite{reed2003expanded} in a glucose limited minimal medium in aerobic conditions with maximum glucose uptake $u=6$mmol/gDWh (the default case provided with the model). After removing leaves this space has dimension $D=233$ for $N=666$ reactions. 
The uniform sampling of the underlying convex polytope has been performed by an hit-and-run Montecarlo Markov chain\cite{Turcin:1971}, whose convergence is guaranteed in feasible times\cite{Lovasz:1999p4121} upon  handling ill-conditioning\cite{uniformell} (that can be severe exactly because of thermodynamic unfeasibilities). Thermodynamically unfeasible cycles have been enumerated for this network\cite{deMartino12} and they can be detected and removed in many ways\cite{schellenberger2011elimination, de2012scalable, saa2016ll}. They have been corrected in a minimal way by  adjusting the gauge parameter $L$ in turn to silence  the less intense reaction flux that amounts to a projection onto  the thermodynamically feasible space\cite{DeMartino:2013p4115}.
Results for the overall distribution of flux instensities $|f_i|$ are shown in Fig 2, before and after correcting for thermodynamic unfeasibilities.
\begin{figure}[h!!!!!!!!!]
\begin{center}
\includegraphics[width=0.55\textwidth]{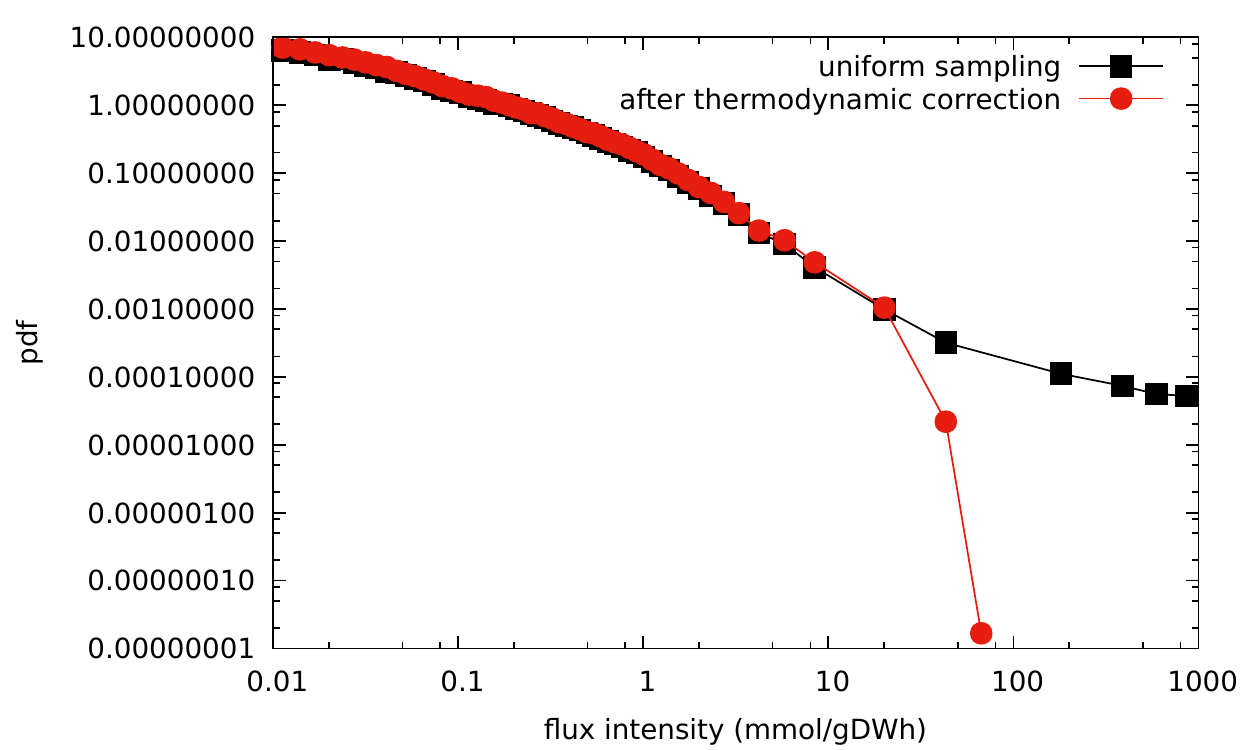}
\includegraphics[width=0.45\textwidth]{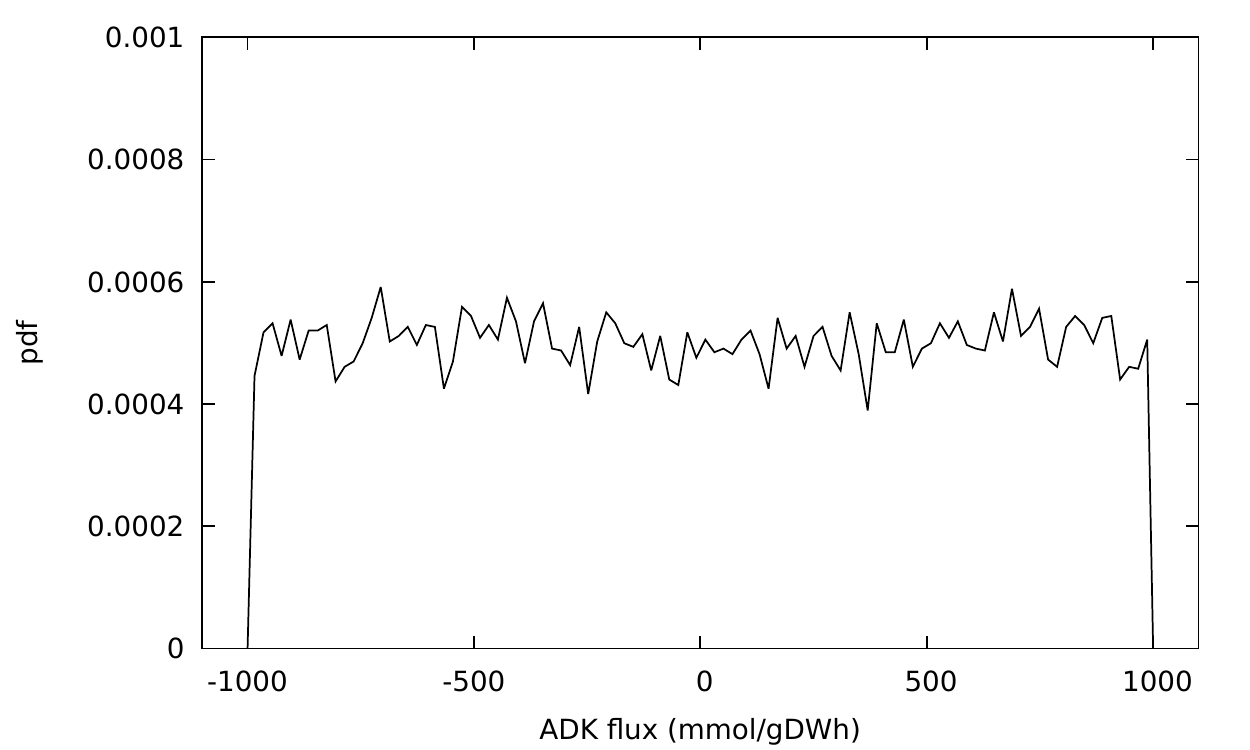}
\includegraphics[width=0.45\textwidth]{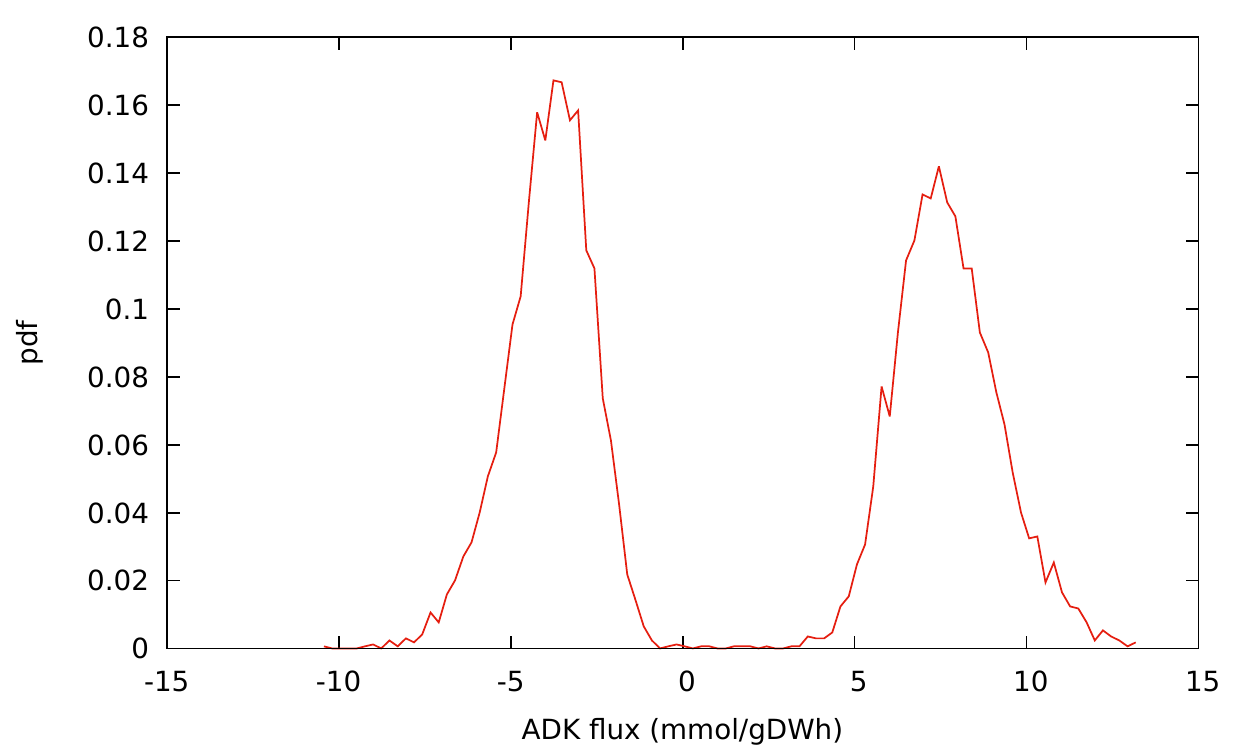}
\end{center}
\caption{Top: Overall flux instensity distribution in log-log scale before and after thermodynamic correction of typical steady states of the model of the metabolism of E.Coli under study from sampling   $R=10^5$ configurations ($N=666$ reactions).  The maximum limiting glucose uptake is $u=6$mmol/gDWh while the arbitrary constant for reaction bounds is $10^3$. Bottom: histogram of Adenylate Kinase before (left) and after thermodynamic correction (right).}
\end{figure}
The long tail corresponding to the uniform sampling depends on the arbitrary constant fixed for flux bounds (in this case $10^3$) and it can be extended thus arbitrarely.
In regard to single reaction flux distributions, we show in Fig 2 (bottom) the retrieved histogram of adenylate kinase, an important enzyme involved in energy homeostasis performing the reaction $2 ADP \leftrightarrow ATP+AMP$. 
Since this reaction is involved in an unifeasible cycle\cite{deMartino12}, its distribution results to be flat and wide upon uniform sampling, while after thermodynamic  correction shows a bimodal fashion opening possible speculations about the presence of two effective states for energetic homeostasis.
In order to keep track in a more rigorous way the scales of the metabolic space in dependence of physical limiting factors we performed a principle component analysis of the sampled configuration upon varying uptake conditions.
The square root of the retrieved  eigenvalues are proportional to the diameters of the maximum inscribed  ellipsoid into the convex hull of the space and thus quantify more rigorously its scales. 
We show in Fig  3  (left) the $100$th largest eigenevalues before and after thermodynamic correction. At odds with the interesting case of assets cross-correlations in financial markets\cite{marsili2002dissecting}, here the largest eigenvalues for the uniform sampling correspond to  unfeasible cycles that desapper after thermodynamic correction. In fig 3 (right) it is shown that the two highest eigenvalues scale linearly with the maximum glucose uptake after taking into account thermodynamics, while they do not in the infeasible case, that is indeed scale free.

\begin{figure}[h!!!!!!!!!]
\begin{center}
\includegraphics[width=0.45\textwidth]{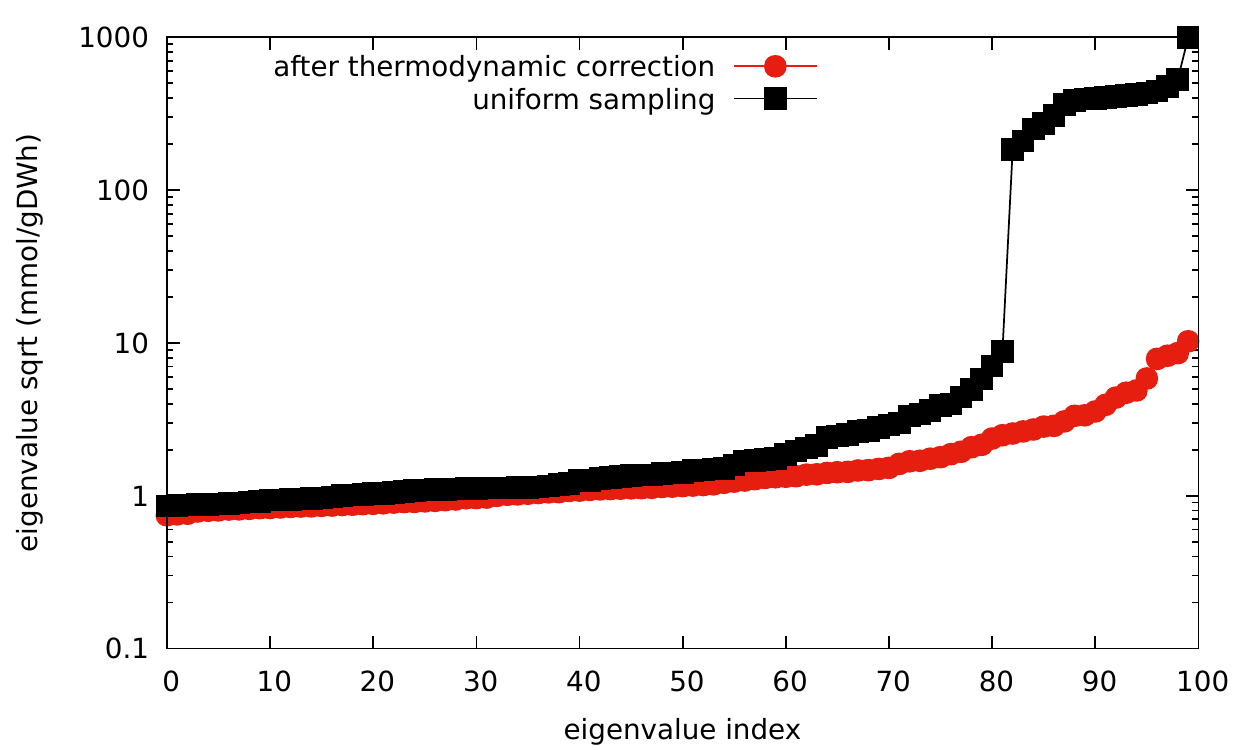}
\includegraphics[width=0.45\textwidth]{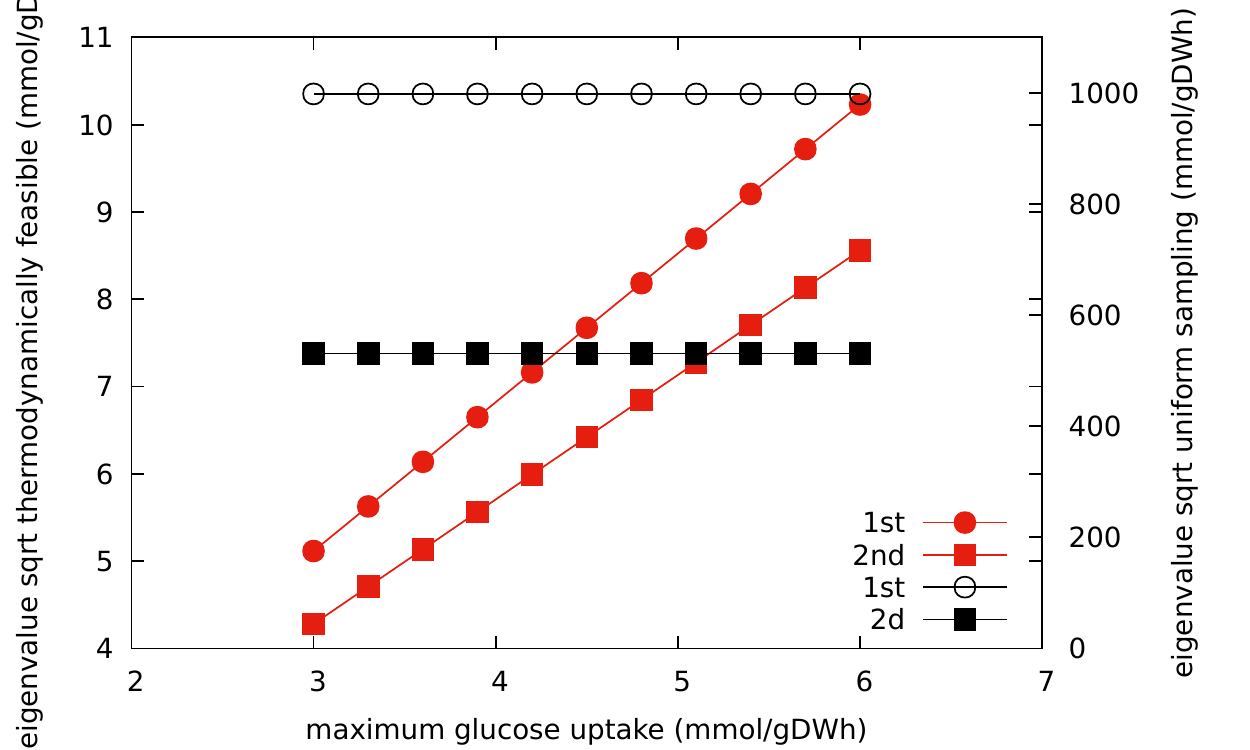}
\end{center}
\caption{Left: The 100th largest eigenvalues (square root) from a principal component analysis of the flux sampling before and after thermodynamics correction. Right: First and second largest eigenvalues (square root) as a function of the maximum glucose uptake before and after thermodynamic correction (notice the different $y$ scales).}
\end{figure}

\subsection*{Towards symmetry breaking in toy models.}
We consider the toy model in Fig 4, where three units uptake some nutrients with intensity $u_i$: they either use it ($y_i$) or exchange it ($x_i$) and resources are globally limited by a certain constant $U$. 
We have the constraints (where we identify $4=1$)
\begin{eqnarray}
u_i\geq 0 \\
\sum_i u_i \leq U \\
u_i = y_i + x_{i+1}-x_i \\
y_i \geq 0
\end{eqnarray}
Because of thermodynamic constraints, the $x_i$ cannot have the same sign.
Suppose  that $\sum_i y_i=\sum u_i$  is maximized, this will be equal to  $U$. This leads to some degeneracy and each configuration $u_i$  has a phase space volume (to which probability is proportional in case of a flat prior)
\begin{equation}
 V(\{u_i\})=\int_{x_{i+1}-x_i\leq u_i} \prod d x_i (\textrm{Not all the same sign})
\end{equation}  
For three nodes we can evaluate the integral dividing it in the $6$ integrals in which the sign $s_i$ of the $x_i$ is fixed (we exclude the $2$ configurations in which signs are equal). For instance consider the integral $I$ upon fixing $s_1=+1$ and $s_2=s_3=-1$.
We replace the $x_i$ with positive definite variables $r_i \geq 0$: $x_i = s_i r_i$. We have the constraints (from $x_{i+1}\leq x_i +u_i$)
\begin{eqnarray}
-r_1-r_2 \leq u_1 \\ 
r_2-r_3 \leq u_2 \\
r_3+r_1\leq u_3
\end{eqnarray}
the first constraint is trivially satisfied, we can integrate $0 \leq r_1 \leq u_3-r_3$ and $0 \leq r_2 \leq u_2+r_3$ obtaining
\begin{equation}
I = \int_0^{u_3} (u_3-r_3)(u_2+r_3) dr_3 = \frac{u_3^2}{2}(u_2+u_3/3)
\end{equation}
Considering the other cases that can be obtained by symmetry
we have finally after some calculations
\begin{equation}
2 V(u_1,u_2,u_3) = U\sum_i u_i^2-\frac{1}{3}\sum_i u_i^3
\end{equation}
This function has a minimum for the symmetric solution $u_i=U/3$ and three (equal) maxima where one unit is uptaking all resources, e.g. $u_1=U$ and $u_2=u_3=0$.
\begin{figure}[h]
  \fbox{\begin{minipage}[]{140pt}
    \includegraphics[height=140pt,width=140pt,angle=270]{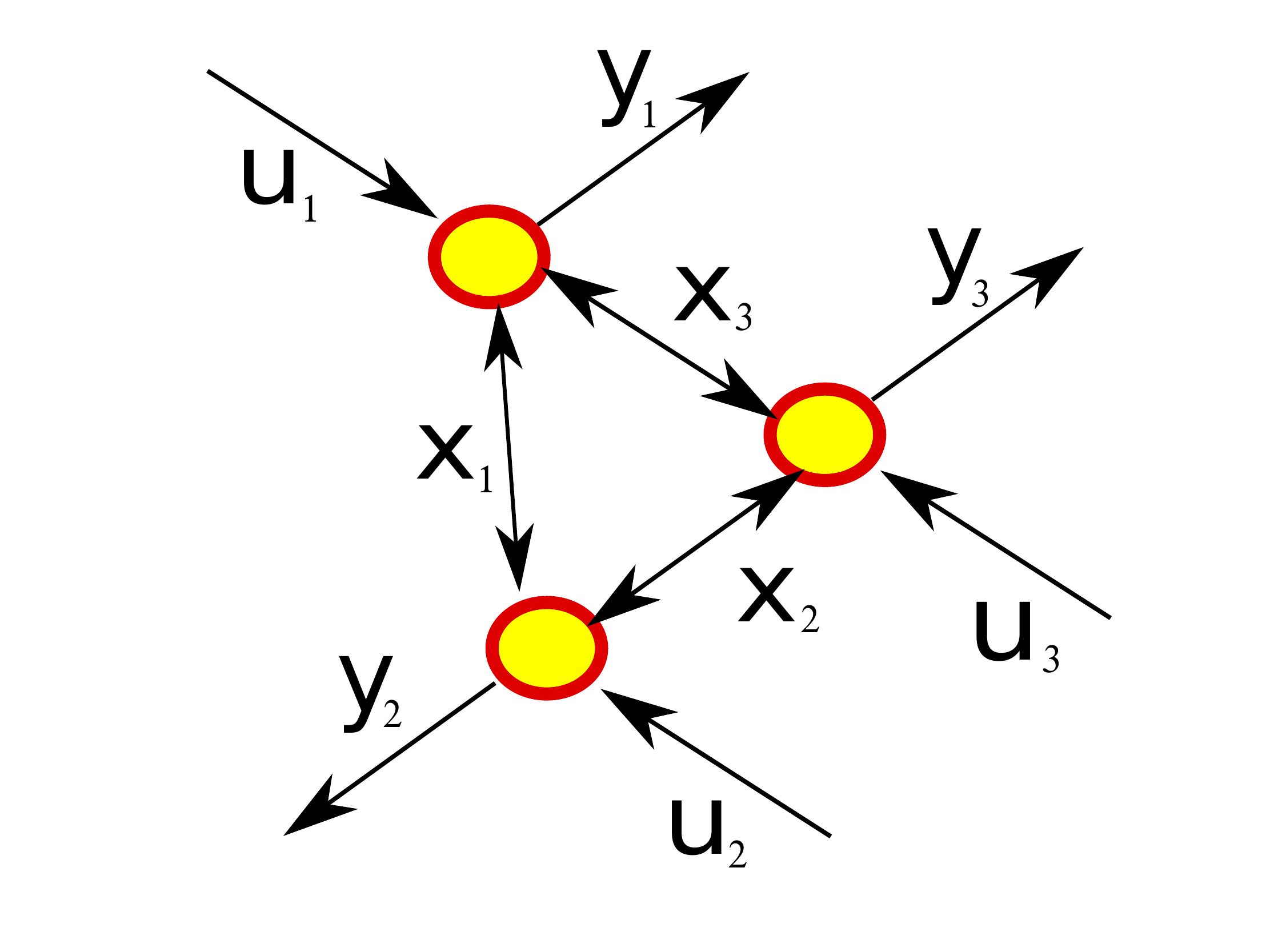}
  \end{minipage}}
  \hfill
  \fbox{\begin{minipage}[]{270pt}
    \includegraphics[height=150pt,width=270pt]{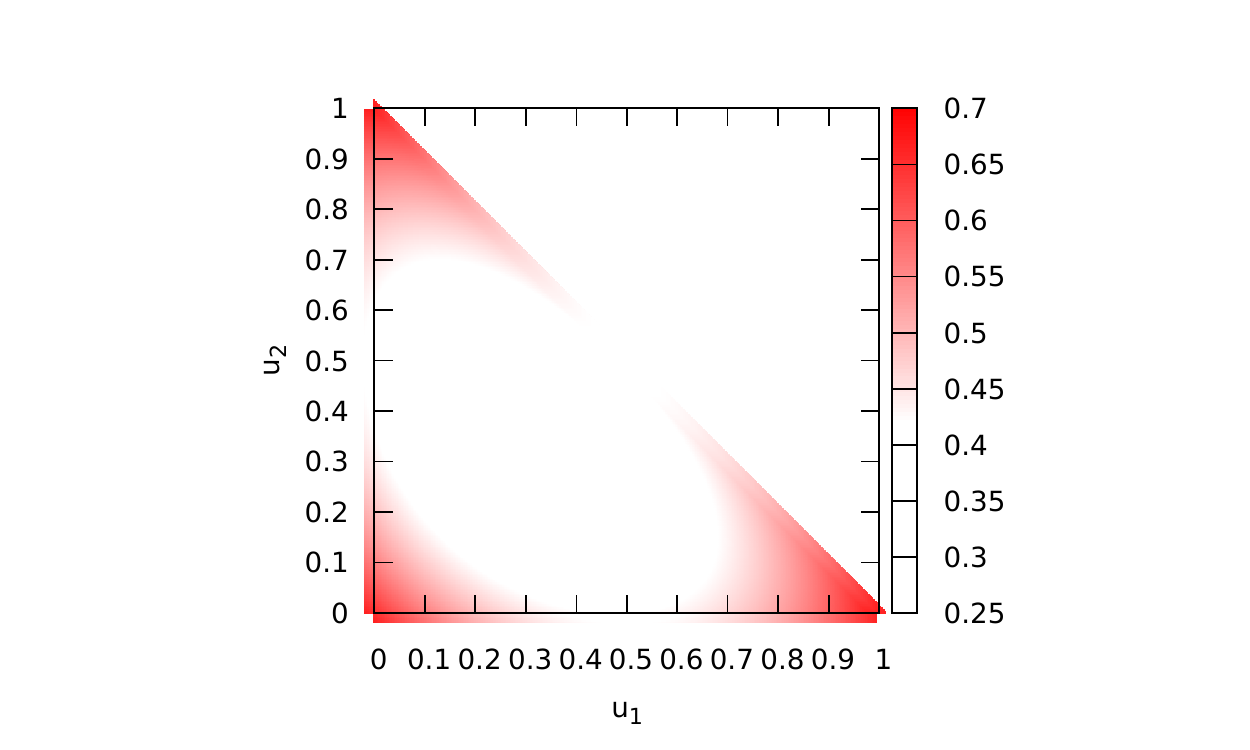}
  \end{minipage}}
  \caption{Left: Toy network model illustrating the mechanism. Right: contour plot of the phase space volume $V$ in the plane $(u_1,u_2)$ when $U=1$}
\end{figure}

\section*{Conclusions}
Although scientists have established  to be impossible under the laws of physics, perpetual motion continues to capture the imagination of inventors and to enter in a subtle way in scientific research as well.
In this work it has been  shown that in order to meaningfully assess scales in metabolic networks  it would be necessary to take into account thermodynamics, given the  free energy-transducing nature of these sytems. It has beeen shown in one realistic example that the long tails inferred in fluxes distributions in metabolic network models are a spurious effect due to reactions involved in infeasible cycles.
From the  positive side, such an energy balance analysis can be rewarding from several aspects, since, apart from assessing metabolites concentrations\cite{Hoppe:2007cr}, free energies\cite{Henry:2007xz}, potentially regulated sites\cite{Kummel:2006kl},
it reveals that the metabolic space has a more complex multimodal structure.
The additional layer of complexity provided by thermodynamics could enrich the possibility of biophysical modelling especially in presence of heterogeneity, cell-to-cell interactions and/or metabolic shuttling, and from a theroretical viewpoint could potentially lead to new variational principles\cite{de2014inferring}.
A simple toy model has been provided to illustrate a possible mechanism of symmetry breaking while maximizing for an objective function.
On the other hand the  inference from  such constraints is challenging from a computational point of view given the non-convexity of the space\cite{reimers2015obstructions}. Suitable extensions of fast approximate analytical methods for the convex case\cite{fernandez2016fast, braunstein2017analytic} could be very promising in this respect.

\section*{Acknowledgments}
The research leading to these results has received funding from the People Programme (Marie Curie Actions) of the European Union's Seventh Framework Programme ($FP7/2007-2013$) under REA grant agreement $n[291734]$.
\section*{References}    
\bibliographystyle{unsrt}
\bibliography{reference}

\providecommand{\noopsort}[1]{}\providecommand{\singleletter}[1]{#1}%
\begin{thebibliography}{10}

\bibitem{cardy1996scaling}
John Cardy.
\newblock {\em Scaling and renormalization in statistical physics}, volume~5.
\newblock Cambridge university press, 1996.

\bibitem{zinn2002quantum}
Jean Zinn-Justin.
\newblock Quantum field theory and critical phenomena.
\newblock 2002.

\bibitem{schneidman2006weak}
Elad Schneidman, Michael~J Berry, Ronen Segev, and William Bialek.
\newblock Weak pairwise correlations imply strongly correlated network states
  in a neural population.
\newblock {\em Nature}, 440(7087):1007--1012, 2006.

\bibitem{mora2011biological}
Thierry Mora and William Bialek.
\newblock Are biological systems poised at criticality?
\newblock {\em Journal of Statistical Physics}, 144(2):268--302, 2011.

\bibitem{levina2007dynamical}
Anna Levina, J~Michael Herrmann, and Theo Geisel.
\newblock Dynamical synapses causing self-organized criticality in neural
  networks.
\newblock {\em Nature physics}, 3(12):857--860, 2007.

\bibitem{mastromatteo2011criticality}
Iacopo Mastromatteo and Matteo Marsili.
\newblock On the criticality of inferred models.
\newblock {\em Journal of Statistical Mechanics: Theory and Experiment},
  2011(10):P10012, 2011.

\bibitem{tkavcik2015thermodynamics}
Ga{\v{s}}per Tka{\v{c}}ik, Thierry Mora, Olivier Marre, Dario Amodei,
  Stephanie~E Palmer, Michael~J Berry, and William Bialek.
\newblock Thermodynamics and signatures of criticality in a network of neurons.
\newblock {\em Proceedings of the National Academy of Sciences},
  112(37):11508--11513, 2015.

\bibitem{lane2015vital}
Nick Lane.
\newblock {\em The vital question: energy, evolution, and the origins of
  complex life}.
\newblock WW Norton \& Company, 2015.

\bibitem{orth2010flux}
Jeffrey~D Orth, Ines Thiele, and Bernhard~{\O} Palsson.
\newblock What is flux balance analysis?
\newblock {\em Nature biotechnology}, 28(3):245--248, 2010.

\bibitem{patil2004use}
Kiran~Raosaheb Patil, Mats {\AA}kesson, and Jens Nielsen.
\newblock Use of genome-scale microbial models for metabolic engineering.
\newblock {\em Current opinion in biotechnology}, 15(1):64--69, 2004.

\bibitem{barabasi2011network}
Albert-L{\'a}szl{\'o} Barab{\'a}si.
\newblock The network takeover.
\newblock {\em Nature Physics}, 8(1):14, 2011.

\bibitem{almaas2004global}
E\_ Almaas, B~Kovacs, T~Vicsek, ZN~Oltvai, and A-L Barab{\'a}si.
\newblock Global organization of metabolic fluxes in the bacterium escherichia
  coli.
\newblock {\em Nature}, 427(6977):839--843, 2004.

\bibitem{monod1949growth}
Jacques Monod.
\newblock The growth of bacterial cultures.
\newblock {\em Annual Reviews in Microbiology}, 3(1):371--394, 1949.

\bibitem{bremer1996modulation}
Hans Bremer and Patrick~P Dennis.
\newblock Modulation of chemical composition and other parameters of the cell
  by growth rate.
\newblock 1996.

\bibitem{beard2002energy}
Daniel~A Beard, Shou-dan Liang, and Hong Qian.
\newblock Energy balance for analysis of complex metabolic networks.
\newblock {\em Biophysical journal}, 83(1):79--86, 2002.

\bibitem{ataman2015heading}
Meric Ataman and Vassily Hatzimanikatis.
\newblock Heading in the right direction: thermodynamics-based network analysis
  and pathway engineering.
\newblock {\em Current Opinion in Biotechnology}, 36:176--182, 2015.

\bibitem{reimers2014metabolic}
Arne~C Reimers.
\newblock {\em Metabolic Networks, Thermodynamic Constraints, and Matroid
  Theory}.
\newblock PhD thesis, Freie Universit{\"a}t Berlin, 2014.

\bibitem{polettini2014irreversible}
Matteo Polettini and Massimiliano Esposito.
\newblock Irreversible thermodynamics of open chemical networks. i. emergent
  cycles and broken conservation laws.
\newblock {\em The Journal of chemical physics}, 141(2):07B610\_1, 2014.

\bibitem{deMartino12}
D~{De Martino}.
\newblock Thermodynamics of biochemical networks and duality theorems.
\newblock {\em Phys. Rev. E}, 87:053108, 2013.

\bibitem{DeMartino:2013p4115}
Daniele {{De Martino}}, Fabrizio Capuani, Matteo Mori, Andrea {{De Martino}},
  and Enzo Marinari.
\newblock Counting and correcting thermodynamically infeasible flux cycles in
  genome-scale metabolic networks.
\newblock {\em Metabolites}, 3(4):946--966, 2013.

\bibitem{reed2003expanded}
Jennifer~L Reed, Thuy~D Vo, Christophe~H Schilling, and Bernhard~O Palsson.
\newblock An expanded genome-scale model of escherichia coli k-12 (i jr904
  gsm/gpr).
\newblock {\em Genome biology}, 4(9):R54, 2003.

\bibitem{Turcin:1971}
V~Turcin.
\newblock On the computation of multidimensional integrals by the {Monte}
  {Carlo} method.
\newblock {\em Th Probab Appl}, 16:720--724, 1971.

\bibitem{Lovasz:1999p4121}
L{\'a}szl{\'o} Lov{\'a}sz.
\newblock Hit-and-run mixes fast.
\newblock {\em Math Program}, 86(3):443--461, 1999.

\bibitem{uniformell}
Daniele De~Martino, Matteo Mori, and Valerio Parisi.
\newblock Uniform sampling of steady states in metabolic networks:
  Heterogeneous scales and rounding.
\newblock {\em PLoS ONE}, 10(4):e0122670, 2015.

\bibitem{schellenberger2011elimination}
Jan Schellenberger, Nathan~E Lewis, and Bernhard~{\O} Palsson.
\newblock Elimination of thermodynamically infeasible loops in steady-state
  metabolic models.
\newblock {\em Biophysical journal}, 100(3):544--553, 2011.

\bibitem{de2012scalable}
Daniele De~Martino, Matteo Figliuzzi, Andrea De~Martino, and Enzo Marinari.
\newblock A scalable algorithm to explore the gibbs energy landscape of
  genome-scale metabolic networks.
\newblock {\em PLoS Comput Biol}, 8(6):e1002562, 2012.

\bibitem{saa2016ll}
Pedro~A Saa and Lars~K Nielsen.
\newblock ll-achrb: a scalable algorithm for sampling the feasible solution
  space of metabolic networks.
\newblock {\em Bioinformatics}, page btw132, 2016.

\bibitem{marsili2002dissecting}
Matteo Marsili et~al.
\newblock Dissecting financial markets: sectors and states.
\newblock {\em Quantitative Finance}, 2(4):297--302, 2002.

\bibitem{Hoppe:2007cr}
Andreas Hoppe, Sabrina Hoffmann, and Hermann Holzhutter.
\newblock Including metabolite concentrations into flux balance analysis:
  thermodynamic realizability as a constraint on flux distributions in
  metabolic networks.
\newblock {\em BMC Systems Biology}, 1(1):23, 2007.

\bibitem{Henry:2007xz}
Christopher Henry, Linda Broadbelt, and Vassily Hatzimanikatis.
\newblock Thermodynamics-based metabolic flux analysis.
\newblock {\em Biophys. J.}, 92(5):1792, 2007.

\bibitem{Kummel:2006kl}
Anne Kummel, Sven Panke, and Matthias Heinemann.
\newblock Putative regulatory sites unraveled by network-embedded thermodynamic
  analysis of metabolome data.
\newblock {\em Molecular Systems Biology}, 2:2006.0034, 2006.

\bibitem{de2014inferring}
Daniele De~Martino, Fabrizio Capuani, and Andrea De~Martino.
\newblock Inferring metabolic phenotypes from the exometabolome through a
  thermodynamic variational principle.
\newblock {\em New Journal of Physics}, 16(11):115018, 2014.

\bibitem{reimers2015obstructions}
Arne~C Reimers.
\newblock Obstructions to sampling qualitative properties.
\newblock {\em PloS one}, 10(8):e0135636, 2015.

\bibitem{fernandez2016fast}
Jorge Fernandez-de Cossio-Diaz and Roberto Mulet.
\newblock Fast inference of ill-posed problems within a convex space.
\newblock {\em Journal of Statistical Mechanics: Theory and Experiment},
  2016(7):073207, 2016.

\bibitem{braunstein2017analytic}
Alfredo Braunstein, Anna~Paola Muntoni, and Andrea Pagnani.
\newblock An analytic approximation of the feasible space of metabolic
  networks.
\newblock {\em arXiv preprint arXiv:1702.05400}, 2017.

\end{thebibliography}




\end{document}